\documentclass[%
 reprint,
 amsmath,amssymb,
 aps,
prb,
]{revtex4-1}

\usepackage{graphicx}% Include figure files
\usepackage{dcolumn}% Align table columns on decimal point
\usepackage{bm}% bold math
\begin{document}

\preprint{APS/123-QED}

\title{Effect of the flash annealing on the impurity distribution and the electronic structure in the inversion layer}% Force line breaks with \\

\author{Tomohiro Sakata}
 \altaffiliation[Also at ]{Materials Science Department, Nara Institute of Science and Technology.}%Lines break automatically or can be forced with \\
\author{Sakura N Takeda}%
\author{Makoto Morita}%
\author{Nur I Ayob}%
\author{Hiroki Tabata}%
\author{Hironori Matsuoka}%
\author{Hiroshi Daimon}
 \email{s-tomohiro@ms.naist.jp}
\affiliation{%
 Graduate School of Materials Science, Nara Institute of Science and Technology, Nara 630-0192, Japan\\
 }%

\date{\today}% It is always \today, today,

\begin{abstract}
Hole subband structure under strong band bending such a Pb on Si(111) and Indium on Si(111) have been investigated by angle-resolved photoelectron spectroscopy(ARPES). Energy levels of hole subband structure which indicate the quantized levels in inversion layer are strongly depend on band bend shape which can be controlled by the impurity concentration of substrate. Meanwhile, the discrepancy for the suband energy separation between experimental results and calculation results is also observed. In this study, we aim to clarify the relationship between flash annealing and impurity concentration and the hole subband. From this results, it was found out that high temperature flash annealing at 1250 $^\circ$C has considerable effect on the impurity concentration at subsurface region by Secondary Ion Mass spectroscopy (SIMS) and our diffusion model. This effect makes the band bend shape and subband energy separation change. Moreover, It was revealed that the reduction of the impurity distribution was inhibited less than 900 $^\circ$C of  the flashing temperature.
%\begin{description}
%\end{description}
\end{abstract}

\maketitle

%\tableofcontents

\section{\label{sec:level1}Introduction}

\begin{figure}[ht]
 \begin{center}
  \includegraphics[width=9cm,clip]{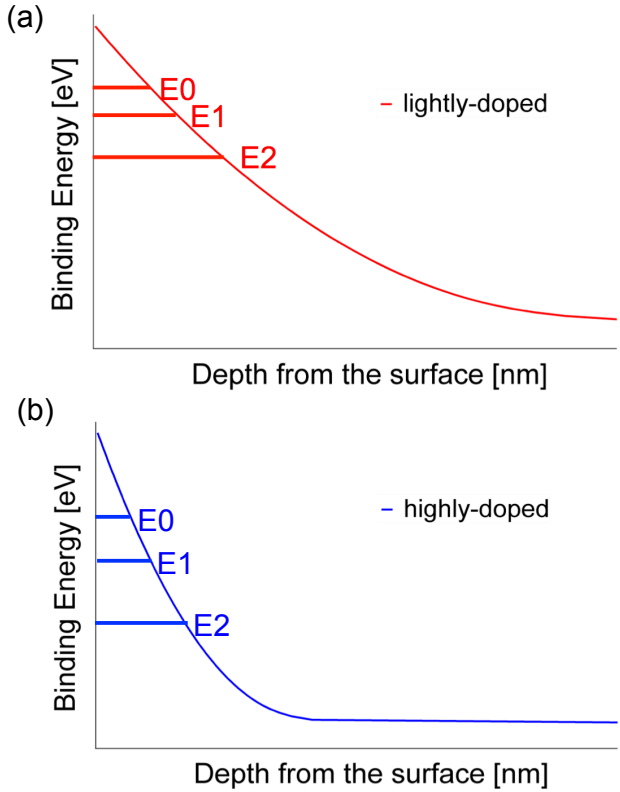}
  \caption{Band bend shape of the lightly-doped sample(a) and the highly-doped sample(b). Subband energy levels are shown in E0, E1, and E2, respectively }
  \label{space}
 \end{center}
\end{figure}

In the reconstruction surface or metal - adsorbed semiconductor surface, excess carrier is accumulated into the electronic surface states. Therefore, in the case of low densities of free carrier like semiconductor, electric charged subsurface region is formed by large screening length. Space charge layer is caused by this phenomenon. Above all, when strong band bending is induced, it has become inversion layer (IL). Electronic structure in IL which formed under the superstructure is already clarified by previous works in the system of In/Si(111)\cite{PhysRevLett.94.037401}\cite{PhysRevLett.82.035318}, Pb/Si(111)\cite{PhysRevLett.94.037401}\cite{PhysRevLett.82.035318}, Pb/Si(001)\cite{e-JSurfSciNanotech}, and PbGa/Si(111)\cite{ASS}. Unlike bulk states, subband structure which indicates electronic structure in IL has quantized levels due to the band bending. Therefore, the narrower energy separations are formed in the wider band bending (shown in  Fig.\ref{space} (a)). On the other hand, the wider energy separation is obtained under stronger band bending due to narrow confined width (shown in  Fig.\ref{space} (b)).The indication of the confined width can be shown in Debye length which can be written by Eq.(\ref{DL}). Therefore this band bend shape depends on the impurity concentration in substrate. The impurity concentration is critical factor for subband energy separations because of changing the confined width. It has already found out that this electronic structure can be written by triangle potential approximation\cite{TPA}. Nevertheless, experimental subband energy separations by ARPES are larger than calculation. 
We assumed that the changing impurity distribution by heat treatment suspected as a major cause of this discrepancy. Heat treatment is essential to get the clean surface. Especially, flash annealing is most effective and simple method for obtaining silicon clean surface. Silicon is heated up to 1250 $^\circ$C and down to RT instantly during flash annealing. On the other hand, this high temperature annealing induces \textit{p}-type band bending on the \textit{n}-type silicon substrate due to carbon \cite{PhysRevB.78.035318}or boron contamination\cite{surfsci1}. It has also possible to change of the impurity distribution by flash annealing because the segregation and reduction of the dopant is caused by long time annealing in vacuum\cite{JAP}\cite{dopant}\cite{PhysRevLett.74.5080}. In this paper, we focused on the effect of the flash annealing on the impurity distribution and clarify the effect of flash annealing on the subband structure by revealing the relationship between flash annealing and the impurity distribution. In addition, we comprehensively understand the band bend shape and subband energy separations with considering flash annealing effect.

\section{\label{sec:level1}Experiment}
We carried out this experiment at a UHV chamber with a  base pressure of less than 5  $\times$ $10^{-10}$ Torr. This equipment was attached with RHEED system, deposition source, an ultra violet source and an electron analyzer (SES2002, VG SCIENTA). The sample is heavily Arsenic-doped Si(111). The resistivity and the impurity concentration are 2 
$\sim$ 2.5 m$\Omega$ cm and 3.5 $\times$ $10^{19}$ atoms/cm, respectively. First each sample was cleaned by flash annealing at 1250 $^\circ$C ($\ge$ 1 sec). The temperature of these samples were raised up to 1250 $^\circ$C within 2 sec from RT and decreased to RT as soon as reaching 1250 $^\circ$C during flash annealing.
In this study, the number of the flash annealing of each sample were strictly controlled so as to see the effect of the flash annealing on the subband energy.
We prepare 8 samples with different number of the flash annealing times which are 0 time, 5 times, 20 times, 40 times, 70 times, 100 times, 250 times and 300 times. After confirming  the Si(111) 7 $\times$ 7 clean surface by RHEED, Pb was deposited on the clean surface for a few ML at RT. 
Then, the sample was annealed at 300 $^\circ$C for few minutes and we confirmed Si(111) $\sqrt{3}$ $\times$ $\sqrt{3}$ - Pb SIC structure\cite{PhysRevB.60.5653}. The band structure of these sample were measured by ARPES (Angle Resolved Photoelectron Spectroscopy). Furthermore the samples which were heated by flash annealing 0 times, 1 time, 5 times, 100 times, and 300 times were measured by SIMS (Secondary Ion Mass Spectroscopy) to see the impurity concentration.

\section{\label{sec:level1}Results and Discussion}
\subsection{\label{sec:level2}Results by ARPES and SIMS}
\begin{figure}[ht]
 \begin{center}
  \includegraphics[width=9cm,clip]{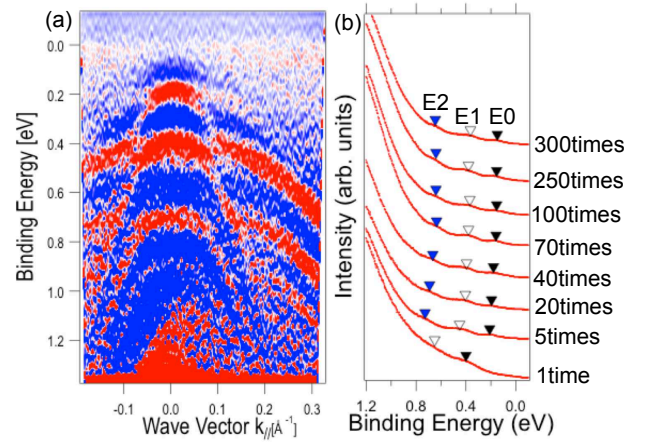}
  \caption{(a) Photoelectron intensity map with 300 times annealing(b)Photoelectron spectra at $\Gamma$ point between 1time and 300times }
  \label{spectrum}
 \end{center}
\end{figure}

The hole subband structures from the sample flash annealed for 300 times and the photoelectron spectra at $\Gamma$ point are shown in Fig.\ref{spectrum} (a), (b) respectively. The high intensity in the photoelectron spectrum is indicated by red color. These subband structures were measured along [11-2]. We can observe the subband energy levels at $\Gamma$ point in the all samples (1 times-300  times). We named these subband energy levels E0, E1, E2 in order of increasing binding energy. All energy levels were determined by Gaussian fitting.
There is a systematic energy shift of peaks E0, E1, E2 depending on the number of times.
We will discuss the dependency later together with the result of SIMS described below.

\begin{figure}[ht]
 \begin{center}
  \includegraphics[width=8cm,clip]{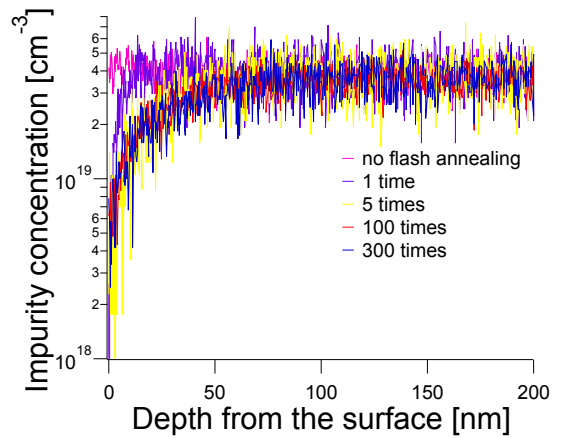}
  \caption{The impurity (Arsenic) distribution which depends on flash annealing times by SIMS}
  \label{Band}
 \end{center}
\end{figure}

The distribution of impurity concentration in the samples which were flash annealed for 1, 5, 100, 300 times and not flash-annealed are investigated by SIMS as shown in Fig.\ref{Band}. The impurity concentration in no flash-annealed sample is constantly distributed from the surface to the bulk. On the other hand, these impurity concentration in the sample with flash annealing decreases  in the subsurface region. This indicates that flash annealing have considerable effect on the impurity distribution curve.

\begin{figure}[ht]
 \begin{center}
  \includegraphics[width=8cm,clip]{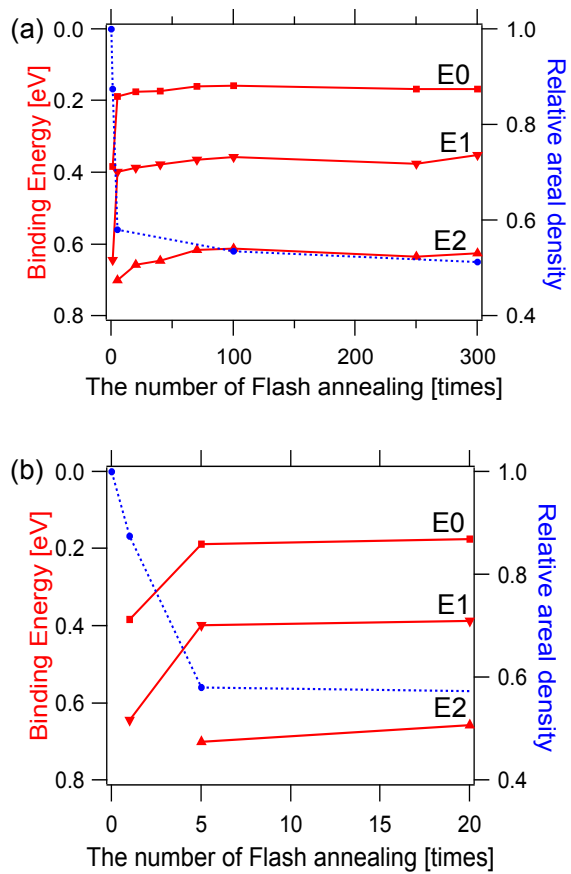}
  \caption{The dependency of the energy separations (solid line, red) and relative areal density (dash line, blue) with flash annealing times between 0times and 300 times(a) and between 0 times and 20 times(b)}
  \label{ESandAD}
 \end{center}
\end{figure}

In order to estimate the amount of the reduction of the impurity at subsurface region quantitatively, the relative arial density of the impurity which shows in Eq.(\ref{rad.}) was introduced. 
\begin{eqnarray}
\frac{\int_0^x C(x, t) dx}{\int_0^x C(x, 0) dx} \label{rad.}
\end{eqnarray}
Here, \textit{x} is depth from the surface and \textit{t} is time which  the sample was at maximum temperature for, respectively. \textit{C(x, t)} is the impurity distribution.
Therefore, the denominator of Eq.(\ref{rad.}) indicates the areal density of the impurity in as- shipped wafer. In Fig.  \ref{ESandAD}, Subband energy levels, E0, E1, and E2 were defined from Fig.  \ref{spectrum} against the number of times flash annealed  and the relative areal density of impurity derived from  Fig. \ref{Band}, respectively. 
In the region in between 0 times and 5 times (region I), the relative areal density was decreased rapidly. The energy levels (E0, E1) shift toward lower binding energy, significantly in this region. The energy shifts of E0 and E1 from 0 times to 5 times were 0.19 eV and 0.26 eV, respectively. The energy separation between E0 and E1 decreases by 20 \symbol{"25}. The decrease of the relative arial density make the slope of the band bend shape slack. Therefore, Energy levels were shifted toward low binding energy and energy separations are decreased.
In the region in between 5 times and 100 times (region I\hspace{-.1em}I), the slight reduction of the relative areal density was observed. Energy levels E0, E1 and E2 also slightly shifted to lower binding energy. The amount of the shift was 0.03 eV, 0.04 eV and 0.09 eV, respectively. Energy separations (E0-E1, E1-E2) show a reduction of 5 \symbol{"25} and 13 \symbol{"25}, respectively. 
In the region between 100 times and 300 times (region I\hspace{-.1em}I\hspace{-.1em}I), the areal density was nearly constant. And the energy shift was not observed. Similarly, energy separations were unchanged. 
This result indicates that the energy levels which locate in higher binding energy were strongly influenced by the decreasing the impurity concentration at subsurface region and significant energy shifts occur. As a result, energy separation was decreased by the reduction of the relative areal density.
Therefore, impurity concentration which depend on flash annealing times correlates with subband energy level and energy separation. Especially, the relative arial density of the impurity is remarkably decreased at the region I (less than 5 times). It makes the subband structure change dramatically.

\subsection{\label{sec:level2}Mechanism of out-diffusion of the impurity in silicon}

 \begin{figure}[ht]
 \begin{center}
  \includegraphics[width=8cm,clip]{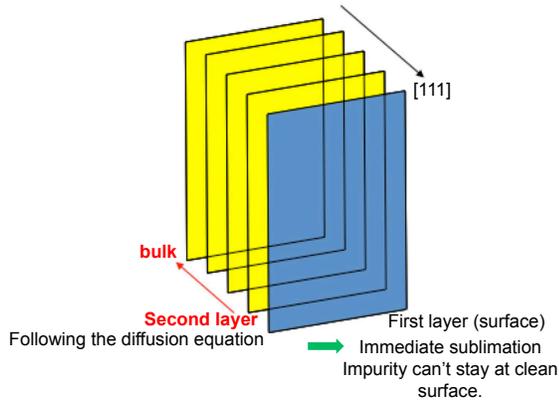}
  \caption{Diffusion model ; impurities from second layer to bulk (shown in yellow area)were diffused by following the diffusion equation. 
  Then, impurities which reached to the clean surface (shown in blue area) were sublimated immediately to the vacuum.}
  \label{model}
 \end{center}
\end{figure}

In this section, in order to reveal the mechanism of the effect of flash annealing on the impurity, we built the impurity diffusion model as shown in  Fig.\ref{model}. In our model, the following is assumed; 

(a) Diffusion condition of the impurity atoms sitting in deeper than second layer from the surface is dominated by diffusion equation which was determined by Fick’s low (\ref{Fick}) and the solution of it (\ref{sol.}) \cite{book1}.
\begin{eqnarray}
\frac{{\partial}C}{{\partial}t}=D\frac{{\partial}^2C}{{\partial}x^2} \label{Fick}
\end{eqnarray}
\begin{eqnarray}
C(x, t)=C_1+C_2 \cdot erfc \frac{x}{2\sqrt{Dt}} \label{sol.}
\end{eqnarray}
Here, C, D is the impurity concentration and diffusion constant respectively. Impurity concentration is represented as the function of time “\textit{t}” and depth from the surface “\textit{x}”. The coefficient $C_1$(=3.5 $\times$ $10^{19}$ $cm^{-3}$) and $C_2$(=-3.5 $\times$ $10^{19}$ $cm^{-3}$) are determined by boundary conditions, C(0, t)=0 $cm^{-3}$ and C($\infty$, 0) =3.5 $\times$ $10^{19}$ $cm^{-3}$. 

(b) All impurity atoms reached to the surface are supposed to sublimate immediately. 

\begin{figure}[!htb]
 \begin{center}
  \includegraphics[width=8cm,clip]{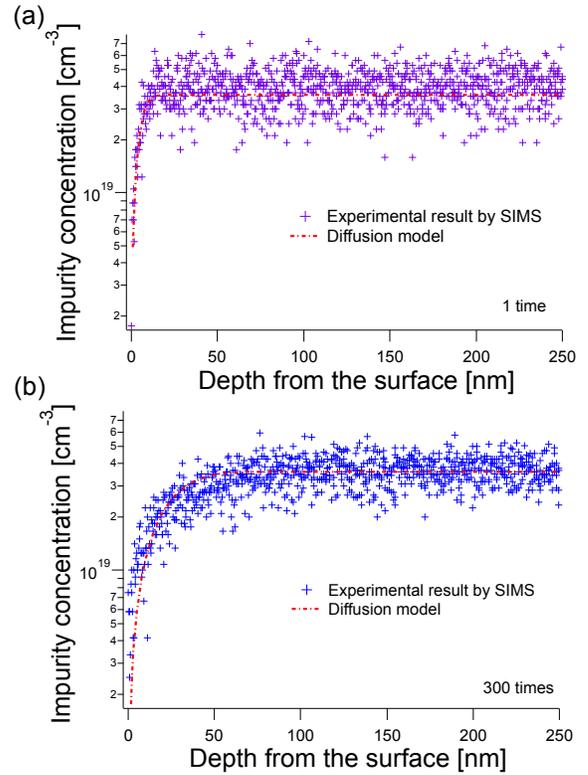}
  \caption{Comparison of impurity distribution curve by SIMS and by calculation based on diffusion model at 1 time(a), and 300 times(b). SIMS result and calculation are indicated by point and dot line, respectively.}
    \label{cal}
 \end{center}
\end{figure}

That is, we assumed that the impurity atoms can not stay at the clean surface for simplicity. The impurity distribution curves were calculated under these condition. D is 2 $\times$ $10^{-13}$ $cm^{2}/sec$ in the case of arsenic in silicon at 1250 $^\circ$C \cite{book4}.
Experimental impurity distribution curve  obtained by SIMS and by calculation using our model are shown in Fig.\ref{cal}. In calculation, we used \textit{t}=0.8 sec for the sample with flash annealed for 1 time and \textit{t}=25 sec for 300 times, respectively. The relation between times and “\textit{t}” wasn't consistent.  This is because the sample temperature was raising continuously during actual flash annealing. On the other hands, “\textit{t}” was defined as the time of 1250 $^\circ$C. The experimental results of the impurity distribution were in good agreement with the calculation by our model.  It indicates by this model that the amount of the impurity reduction is saturated with increasing annealing time. Out of diffusion of the impurity at the subsurface was caused by flash annealing. 

\subsection{\label{sec:level2}Comparison with TPA calculation}
\begin{figure}[h]
 \begin{center}
  \includegraphics[width=8cm,clip]{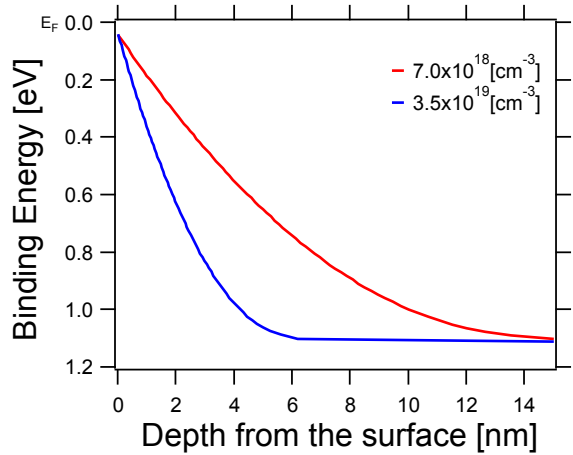}
  \caption{Band bend shape after 300 times flash annealing(red line) and before flash annealing(blue line).}
  \label{bandbending}
 \end{center}
\end{figure}

We calculated the band bending and subband energy separations by considering the decreasing of the impurity in the subsurface region of the sample flash annealing for 300 times. The band bending curve was obtained as the solution of Poisson equation which includes in Debye length\cite{book3}.  The term of impurity concentration which shows in $(n_b+p_b)$ is included in Debye length which is defined by Eq.(\ref{DL})
\begin{eqnarray}
L_D= \sqrt\frac{k_BT\varepsilon}{(n_b+p_b)q^{2}} \label{DL}
\end{eqnarray}
Here $k_B$, \textit{T} and $\varepsilon$ are Bolzman constant, temperature and permittivity, respectively.
Dopant concentration in the subsurface region of the sample with 300 times flash annealing was determined by averaging  the concentration obtained by the SIMS in the region to 20 nm from the surface.  Comparison of the band bending between before flash annealing and after flash annealing (flash annealing times; 300 times) was shown in Fig.\ref{bandbending}. Debye length of the sample  after 300 times flash annealing  was obtained to be 1.57 nm by Eq.(\ref{DL}). On the other hand, Debye length of the sample with original impurity concentration, (=3.5 $\times$ $10^{19}$ $cm^{-3}$) was 0.65 nm. Therefore confined width which subband energy separations depend on becomes 2.4 times wider than original dopant concentration  by flash annealing due to the reduction of impurity.

\begin{table*}
\caption{\label{tab:table3}TPA Calculation results of subband energy separation before flash annealing (original impurity) and after flash annealing; Experimental energy separations are found 0.18 eV (E0-E1) and 0.27 eV (E1-E2) by ARPES}
\begin{ruledtabular}
\begin{tabular}{ccccc}

Flash annealing effect&Original impurity&After flash annealing\\ \hline
  Impurity concentration&3.5 $\times$ $10^{19}$ [$cm^{-3}$]&7.0 $\times$ $10^{18}$ [$cm^{-3}$] \\
E0'-E1' [eV]&0.32 [eV] & 0.18 [eV]    \\   
 E1'-E2' [eV]&0.50 [eV]  & 0.28 [eV] \\
 \end{tabular}
\end{ruledtabular}
\end{table*}

In order to calculate the subband energy levels, we used TPA (; Triangle Potential Approximation) calculation. It was reported by previous work that subband energy levels are in good agreement with the calculated levels by TPA \cite{PhysRevLett.82.035318}. The calculated results with and without considering the reduction of the impurity are compared in Table I. The calculated energy separations between E0’ and E1’, and  E1’ and E2’ after flash annealing were 0.18 eV, 0.28 eV, respectively. On the other hand, the experimental separations obtained by ARPES are 0.18 eV (E0-E1) and 0.27 eV (E1-E2). Unlike calculated energy separations before flash annealing, the subband energy separations after flash annealing by calculation are in good agreement with the experimental one .  Without considering the decrease of the impurity, these subband energy separations are 2 times wider than the experimental one because of narrow Debye length. This result indicates that in the case of using flash annealing method, the shape of the space charge layer at subsurface of silicon is wider than we expect. 

\subsection{\label{sec:level2}Solution}
\begin{figure}[h]
 \begin{center}
  \includegraphics[width=8cm,clip]{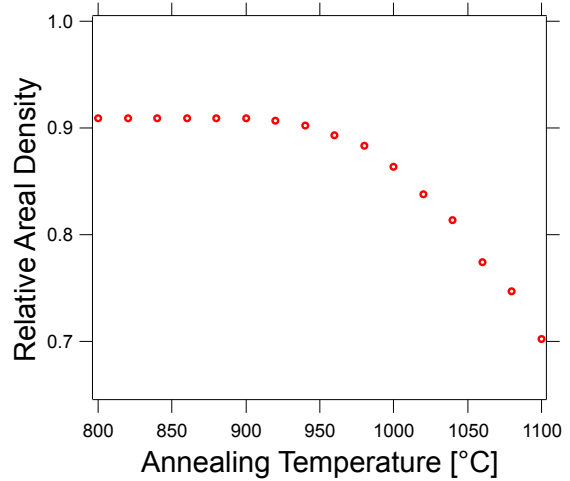}
  \caption{Annealing temperature- dependence of relative areal density which based on diffusion model by calculation}
  \label{temperature}
 \end{center}
\end{figure}

It was revealed that the reduction of the impurity concentration at subsurface region is caused by flash annealing.  Furthermore, the electronic states at subsurface region is dramatically changed by this phenomenon. Therefore, low temperature annealing strategy for getting clean surface was required. However, high quality clean surface is obtained by higher annealing temperature. In this chapter, the solution for avoiding the reduction of the impurity at subsurface is proposed by our diffusion model. Relative areal density by annealing temperature was shown in Fig.\ref{temperature}. Here, Relative areal density is determined by Eq.(\ref{rad.}). Annealing temperature, \textit{t} is 3 sec.  It was suggested by this result that relative areal density is constant below 900 $^\circ$C. In other words, Reduction of the impurity at subsurface was inhibited  in the case of annealing at 900 $^\circ$C.

\section{\label{sec:level1}Conclusion}

In this work, we demonstrated that the impurity concentration decreases at subsurface region  due to the  flash annealing which is conventional method for silicon to make clean surface. The impurity concentration significantly decreased with flash annealing only for 0 times to 5 times and it is constant between 100 times and 300 times. This phenomenon can be described by the thermal diffusion in bulk and the sublimation in the surface. Also, it has the effect to extend the confined width of the space charge layer. Consequently, the energy separations of quantized levels were narrower than we expected. The band bending shape is essential to understand the property of the surface. However, flash annealing influence this shape. Therefore, we should consider the considerable effect of the flash annealing on the impurity distribution. So, there is the need to clean the surface of silicon by low temperature annealing. In our model, it is suggested that the best annealing temperature is approximate 900 $^\circ$C. It was strongly recommended that silicon clean surface was obtained by using wet chemical etching method\cite{surfsci1}\cite{Electrochem}. However, it was suggested that the density of the contamination on the clean surface which was obtained by this wet method is different from it on the cleaned surface by flash annealing because surface contamination is removed by very high temperature of the flash annealing. Therefore, it is necessary to study about the different of the surface property by different method.

\bibliography{apssamp12}% Produces the bibliography via BibTeX.

\end{document}